\documentstyle[psfig,12pt]{article}

\def\lsim{\raise0.3ex\hbox{$\;<$\kern-0.75em\raise-1.1ex\hbox{$\sim\;$}}}
\def\gsim{\raise0.3ex\hbox{$\;>$\kern-0.75em\raise-1.1ex\hbox{$\sim\;$}}}
\def\lsim{\:\raisebox{-0.5ex}{$\stackrel{\textstyle<}{\sim}$}\:}
\def\gsim{\:\raisebox{-0.5ex}{$\stackrel{\textstyle>}{\sim}$}\:}
\def\half{{\textstyle{1 \over 2}}}
\def\fourth{{\textstyle{1 \over 4}}}
\def\etal{{\it et al.}}
\def\ie{{\it i.e.}}

\begin{document}
\begin{titlepage}
\begin{flushright}
HEP-FSU-990520\\
hep-ph/9905422\\
May 1999\\
\end{flushright}
\vspace*{5mm}
\begin{center} 
{\Large \bf The Decay $b\rightarrow s\gamma$, the Higgs Boson Mass, and 
            Yukawa Unification without R-Parity
\footnote{Based on talk given at the SUSY--98 Conference, 11-17 July 1998, 
Oxford, England.}
}\\[15mm]

{\large Marco Aurelio D\'\i az}
\\
\hspace{3cm}\\
{\small       High Energy Group, Department of Physics,
              Florida State University}\\
{\small       Tallahassee, FL 32306-4350, USA}\\
\end{center}
\vspace{5mm}
\begin{abstract}

We review some properties of Bilinear R--Parity Violating models: simple
extensions of the Minimal Supersymmetric Standard Model motivated by
spontaneous breaking of R--Parity. We concentrate on the relaxation of
the bounds on the charged Higgs mass imposed by the measurement of
$B(b\rightarrow s\gamma)$, the effect on the mass of the lightest neutral
Higgs boson, the radiative breaking of the electroweak symmetry, 
the unification of bottom and tau Yukawa coupling, and the relation
of these phenomena to the radiatively generated tau--neutrino mass.

\end{abstract}

\vfill

\end{titlepage}

\setcounter{page}{1}

\section{Introduction}

Explicit Bilinear R--Parity Violation (BRpV) is a simple extension of
the Minimal Supersymmetric Standard Model (MSSM) which has the attractive 
feature of generating neutrino masses radiatively, thus naturally small. 
The origin of the neutrino mass is then linked to supersymmetry 
\cite{SUSYorig}
through the mixing of neutral higgsinos and gauginos with the neutrino.
The study of models which include BRpV terms, and not trilinear (TRpV), 
is motivated by spontaneous R--Parity breaking \cite{spont}, 
where only BRpV terms are generated in the superpotential. The simplest 
model includes violation of R-Parity and lepton number only in the tau 
sector, and the superpotential is
\begin{equation} 
W=h_t\widehat Q_3\widehat U_3\widehat H_2
+h_b\widehat Q_3\widehat D_3\widehat H_1
+h_{\tau}\widehat L_3\widehat R_3\widehat H_1
-\mu\widehat H_1\widehat H_2
+\epsilon_3\widehat L_3\widehat H_2\,,
\label{Wsuppot}
\end{equation}
where $\epsilon_3$ has units of mass and is the only extra term compared 
with the MSSM. The presence of this term induces a non--zero vacuum
expectation value $v_3$ of the tau--sneutrino 
\cite{CGJRV,e3others1,e3others2}. The study of different 
aspects of the phenomenology of this model may be simpler in different 
basis \cite{Ferrandis}. Besides the original one in eq.~(\ref{Wsuppot}), 
a useful basis is defined by the rotation 
$\mu'\widehat H_1'=\mu\widehat H_1-\epsilon_3\widehat L_3$ and
$\mu'\widehat L_3'=\epsilon_3\widehat H_1+\mu\widehat L_3$, where
$\mu'^2=\mu^2+\epsilon_3^2$. The main feature of this basis is that BRpV
is removed from the superpotential. Indeed, the superpotential in the 
rotated basis is given by
\begin{equation} 
W=h_t\widehat Q_3\widehat U_3\widehat H_2
+h_b{{\mu}\over{\mu'}}\widehat Q_3\widehat D_3\widehat H'_1
+h_{\tau}\widehat L'_3\widehat R_3\widehat H'_1
-\mu'\widehat H'_1\widehat H_2
+h_b{{\epsilon_3}\over{\mu'}}\widehat Q_3\widehat D_3\widehat L'_3
\,,\label{WsuppotP}
\end{equation}
with R--Parity non--conservation present in the form of TRpV, with an 
equivalent $\lambda$--coupling given by 
$\lambda'_{333}\equiv h_b\epsilon_3/\mu'$. 

Although BRpV is removed from the superpotential, it is reintroduced in
the soft terms. The relevant terms in the original basis are
\begin{equation}
V_{soft}=m_{H_1}^2|H_1|^2+M_{L_3}^2|\widetilde L_3|^2
-\left[B\mu H_1H_2-B_3\epsilon_3\widetilde L_3H_2+h.c.\right]+...
\label{SoftUnrot}
\end{equation}
After performing the rotation described above, the soft lagrangian 
becomes
\begin{equation}
V_{soft}=m'^2_{H_1}|H'_1|^2+M'^2_{L_3}|\widetilde L'_3|^2-
\bigg[B'\mu'H'_1H_2
-{{\epsilon_3\mu}\over{\mu'^2}}\Delta m^2\widetilde L'_3H'_1
-{{\epsilon_3\mu}\over{\mu'}}\Delta B\widetilde L'_3H_2+h.c.\bigg]+...
\label{SoftRotated}
\end{equation}
where we have defined
$m'^2_{H_1}=(m_{H_1}^2\mu^2+M_{L_3}^2\epsilon_3^2)/\mu'^2$ and 
$M'^2_{L_3}=(m_{H_1}^2\epsilon_3^2+M_{L_3}^2\mu^2)/\mu'^2$ as the new scalar
masses, and $B'=(B\mu^2+B_2\epsilon_3^2)/\mu'^2$ as the new $B$--term. The
last two terms, where $\Delta m^2\equiv m_{H_1}^2-M_{L_3}^2$ and
$\Delta B\equiv B_3-B$, violate R--Parity bilinearly. It is clear that these 
two terms induce a non--zero vev for the rotated tau--sneutrino field
$v'_3\equiv \epsilon_3 v_1+\mu v_3$. In models with universality of soft
terms, the vev $v'_3$ is small because $\Delta m^2$ and $\Delta B$ are
radiatively generated at the weak scale and proportional to the bottom
quark Yukawa coupling. In this case, using the tadpole equations 
\cite{SugraBRpV}, $v'_3$ can be approximated by
\begin{equation}
v'_3\approx -{{\epsilon_3\mu}\over{\mu'^2m_{\tilde\nu^0_{\tau}}^2}}
\left(v'_1\Delta m^2+\mu'v_2\Delta B\right)
\label{App_v3p}
\end{equation}
where we have introduced
\begin{equation}
m_{\tilde\nu^0_{\tau}}^2\equiv {{m_{H_1}^2\epsilon_3^2+M_{L_3}^2\mu^2}
\over{\mu'^2}}+{\textstyle{1\over8}}(g^2+g'^2)(v'^2_1-v_2^2)\,.
\label{sneumassMSSM}
\end{equation}
This mass reduces to the tau--sneutrino mass of the MSSM in the 
$\epsilon_3\rightarrow 0$ limit.

\begin{figure}[htb]
\centerline{\psfig{file=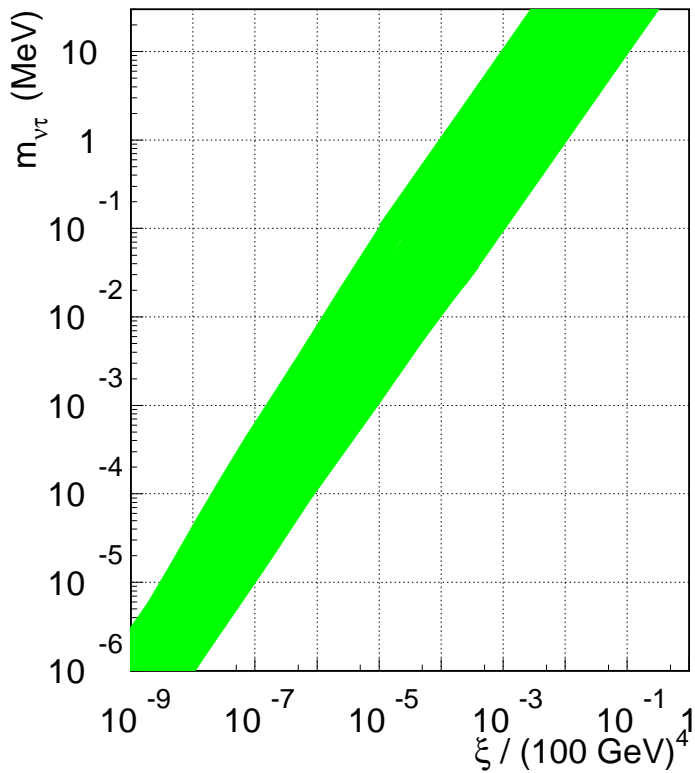,height=11cm}}
\caption{
Tau--neutrino mass $m_{\nu_{\tau}}$ as a function of the parameter
$\xi\equiv v'^2_3\mu'^2$, where $v'_3$ is the tau--sneutrino vacuum 
expectation value in the rotated basis.
} 
\label{mnt_xi_new}
\end{figure} 
The tau--neutrino acquires a mass because it mixes with the neutralinos.
In a see--saw type of mechanism, with the neutralino masses playing the
role of a high scale and $v'_3$ as the low scale, the tau--neutrino mass
is approximately given by the expression
\begin{equation}
m_{\nu_{\tau}}\approx-{{(g^2M'+g'^2M)\mu'^2v'^2_3}\over{
4MM'\mu'^2-2(g^2M'+g'^2M)v'_1v_2\mu'}}
\label{mNeutrinoApp}
\end{equation}
which is naturally small because of eq.~(\ref{App_v3p}). The tau--neutrino 
mass depends strongly on the 
tau--sneutrino vev $v'_3$ as it can be appreciated from 
Fig.~\ref{mnt_xi_new}. In this figure we plot $m_{\nu_{\tau}}$ as a 
function of the parameter $\xi\equiv(\epsilon_3 v_1+\mu v_3)^2=v'^2_3\mu'^2$.
We easily find solutions with neutrino masses from the collider limit
of 17 MeV down to eV. The width of the band in Fig.~\ref{mnt_xi_new} is
related to the parameter $\tan\beta=v_2/v_1$, with the smaller (larger)
values of $\tan\beta$ concentrated at the left (right) of the band.

\section{Unconstrained MSSM--BRpV and $B(b\rightarrow s\gamma)$}

By unconstrained MSSM--BRpV we understand the model where all soft 
parameters are independent at the weak scale, \ie, not embedded into
supergravity. We study the predictions of this model on the branching
ratio $B(b\rightarrow s\gamma)$ varying randomly the soft parameters
at the weak scale \cite{DTV}.

It is well known that the decay $b\rightarrow s\gamma$ is sensible to
physics beyond the Standard Model (SM). The reason is that this decay
is forbidden at tree level, and one--loop contributions from new physics
compete with the SM contribution itself. The theoretical prediction of
the decay rate in the SM, where $W$--bosons and top quarks contribute in
the loops, is $B(b\rightarrow s\gamma)=(3.28\pm0.33)\times10^{-4}$ 
\cite{CMisiakM} including NLO QCD corrections 
\cite{CMisiakM,QCDothers1,QCDothers2}.
This prediction is in agreement at the $2\sigma$ level with the CLEO
official measurement $B(b \to s\gamma)=(2.32\pm 0.57\pm 0.35)\times 10^{-4}$
\cite{CLEO}. Conservatively, this measurement implies
$1.0\times 10^{-4}<B(b \to s\gamma)<4.2\times 10^{-4}$ at $95\%$ C.L.,
which has been modified by the preliminary measurement
$2.0\times 10^{-4}<B(b \to s\gamma)<4.5\times 10^{-4}$ at $95\%$ C.L. 
reported in \cite{CLEO2} after including more data.

\begin{figure}[htb]
\centerline{\psfig{file=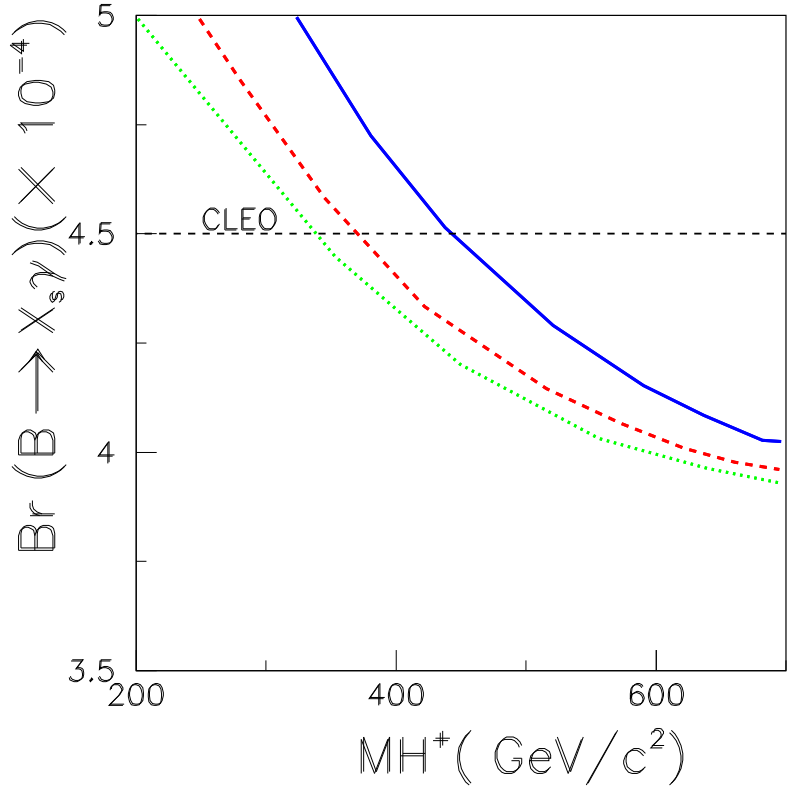,height=11cm}}
\caption{
Lower limit of the branching ratio $B(b\rightarrow s\gamma)$ as a 
function of the charged Higgs mass $m_{H^{\pm}}$ in the limit of very
heavy squark masses. In solid is the MSSM and the other two curves 
correspond to BRpV. The horizontal line corresponds to the upper 
experimental limit from CLEO.
} 
\label{brmh}
\end{figure} 
In two Higgs doublets models of type II, where one Higgs doublet gives 
mass to the up--type fermions and the other to the down--type fermions, 
strong constraints on the charged Higgs mass are obtained because the 
$H^{\pm}$ contribution adds to the $W^{\pm}$ contribution to
$B(b\rightarrow s\gamma)$ \cite{2HDM}. These strong constraints on 
$m_{H^{\pm}}$ are also valid in supersymmetric (SUSY) models with large 
superpartners masses, although relaxed at high values of $\tan\beta$ 
\cite{bsgDiaz1} due to large corrections to the charged Higgs coupling 
to quarks \cite{HudDiaz}. 

In SUSY models with light superpartners the strong constraints on the
charged Higgs mass are no longer valid after the inclusion of chargino,
neutralino, and gluino loops along with squarks 
\cite{BBMR,SUSY1,SUSY2,bsgLater1,bsgLater2,BaerEtal}. 
This is specially due to the chargino 
loops which can cancel completely the charged Higgs loop. Supergravity 
models (SUGRA), with universality of soft mass parameters at the 
unification scale and with radiative breaking of the electroweak symmetry, 
are more constrained ruling out most of the parameter space if $\mu<0$ 
and large $\tan\beta$ \cite{BaerEtal}.

It has been shown that Bilinear R--Parity Violation can relax the bounds
on the charged Higgs mass \cite{DTV}. In this model new particles 
contribute in the loops to $B(b\rightarrow s\gamma)$. Charginos mix
with the tau lepton \footnote{
This mixing is not in conflict with the well measured tau couplings to 
gauge bosons \cite{ADV}.
}, 
therefore, the tau lepton contribute to the decay rate together with 
up-type squarks in the loops. Nevertheless, it was demonstrated that the 
tau contribution can be neglected \cite{DTV}. In a similar way, the 
charged Higgs boson mixes with the two staus \cite{ChaHiggs} forming a 
set of four charged scalars, one of them being the charged Goldstone 
boson. In this way, the staus contribute to the decay rate together with 
up-type quarks in the loops. 

The four charged scalars in the original basis are ${\bf\Phi}^{\pm}=
(H_1^{\pm},H_2^{\pm},\tilde\tau_L^{\pm},\tilde\tau_R^{\pm})$ and the
corresponding mass matrix is diagonalized after the rotation
${\bf S}^{\pm}={\bf R}_{S^{\pm}}{\bf\Phi}^{\pm}$ where ${\bf S}^{\pm}_i$,
$i=1,2,3,4$ are the mass eigenstates (one of them the unphysical Goldstone 
boson). One of the massive charged scalars has similar properties to
the charged Higgs of the MSSM. In BRpV we call the ``charged Higgs boson''
to the charged scalar whose couplings to quarks are larger, \ie, 
maximum $({\bf R}_{S^{\pm}}^{i1})^2+({\bf R}_{S^{\pm}}^{i2})^2$.
Nevertheless, for comparison we have also study the case in which the
``charged Higgs boson'' corresponds to the charged scalar with largest 
components to the rotated Higgs fields $H'^{\pm}_1$ and $H'^{\pm}_2$, 
\ie, maximum $({\bf R'}_{S^{\pm}}^{i1})^2+({\bf R'}_{S^{\pm}}^{i2})^2$.

We neglect in this calculation the contribution of neutralinos, because it 
is small \cite{BBMR}, and that of the gluino whose different squark 
contributions tend to cancel with each other \cite{BaerEtal}. In addition, 
if gaugino masses are universal at the GUT scale, gluinos must be rather 
heavy considering the bound on the chargino mass from LEP2 \cite{DiazKing},
which makes the contribution smaller. We ignore the light gluino window
\cite{Clavelli} because it is inconsistent with the experimental bound on
the mass of the lightest Higgs boson in the MSSM \cite{Diazgluino}.

\begin{figure}[htb]
\centerline{\psfig{file=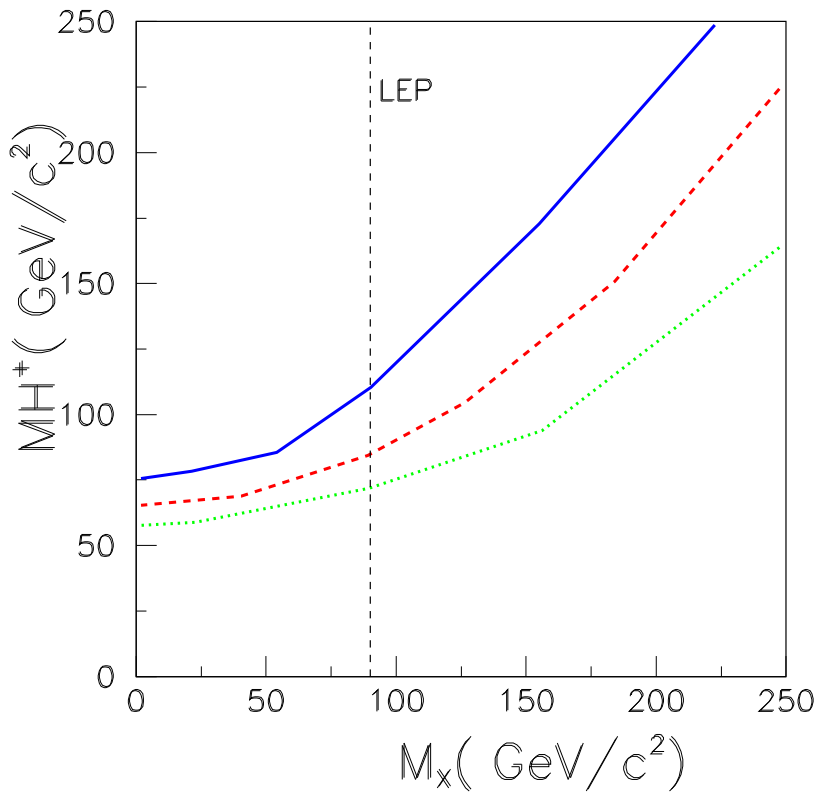,height=11cm}}
\caption{
Lower limit of the charged Higgs mass $m_{H^{\pm}}$ as a function of the 
lightest chargino mass $m_{\chi^{\pm}_1}$ compatible with the CLEO 
measurement. In solid is the MSSM and the other two curves correspond to 
BRpV. The vertical line is the experimental lower limit on 
$m_{\chi^{\pm}_1}$ from LEP.
} 
\label{mhcha}
\end{figure} 
In the limit of very heavy squarks, the strong constraints imposed on the 
charged Higgs mass of the MSSM are relaxed in the MSSM--BRpV as can be
appreciated in Fig.~\ref{brmh}. Above and to the right of the solid line 
are the solutions of the MSSM consistent with the CLEO measurement of 
$B(b\rightarrow s\gamma)$. Without considering theoretical uncertainties, 
the limit on the charged Higgs mass is $m_{H^{\pm}}>440$ GeV. This bound 
is relaxed by about 70 to 100 GeV in BRpV as can be seen from the dotted 
and dashed lines. If a 10\% theoretical uncertainty is considered, the 
MSSM bound reduces to $m_{H^{\pm}}>320$ GeV, but the BRpV bound decreased 
as well such that the reduction of the bound is maintained. The dotted
line corresponds to the charged Higgs with largest couplings to quarks,
a definition that makes more sense in our calculation. The dashed line 
corresponds to a charged Higgs defined by maximum component along 
$H'^{\pm}_1$ and $H'^{\pm}_2$.

Another interesting case is the region of parameter space where the 
charged Higgs and the charginos are light. The limits on light 
$m_{H^{\pm}}$ in the MSSM are also relaxed in BRpV as shown in 
Fig.~\ref{mhcha}.
Solutions consistent with the CLEO measurement of $B(b\rightarrow s\gamma)$
in the MSSM lie over and to the left of the solid curve, implying that
a chargino heavier that 90 GeV requires a charged Higgs heavier than
110 GeV. This bound on $m_{H^{\pm}}$ is relaxed in BRpV by 25 to 35 GeV
as showed by the other two curves. In particular, the charged Higgs can
be lighter than $m_W$ and observable at LEPII. 

\begin{figure}[htb]
\centerline{\psfig{file=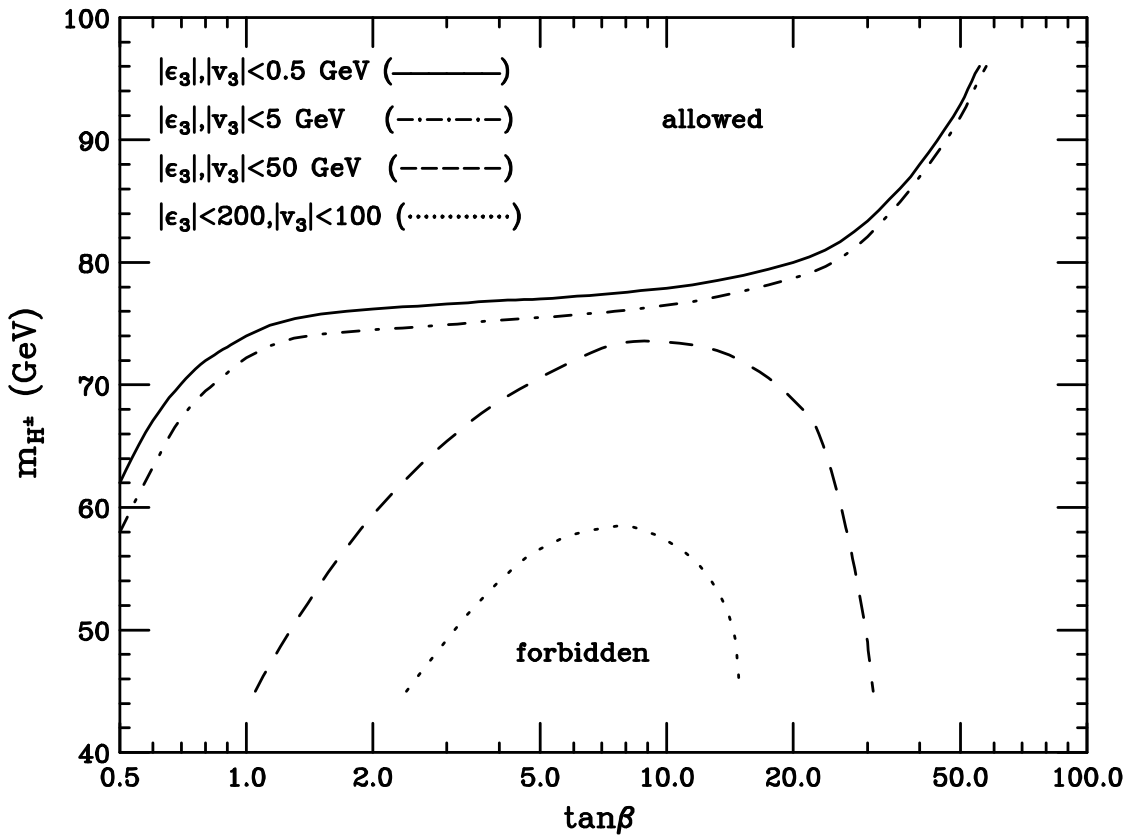,height=11cm}}
\caption{
Lower limit of the charged Higgs mass $m_{H^{\pm}}$ as a function of the 
} 
\label{mchtb}
\end{figure} 
In the MSSM, the charged Higgs can be lighter than $m_W$ after the 
inclusion of radiative corrections \cite{DiazHaberi} in some corners of 
parameter space. In BRpV this situation is not so rare \cite{ChaHiggs}.
In Fig.~\ref{mchtb} we show that for moderate values of $\epsilon_3$
the charged Higgs mass may be lower that $m_W$. Nevertheless, to the 
normal MSSM decay modes we need to add the R-Parity violating decay
modes $H^{\pm}\rightarrow\tau^{\pm}\tilde\chi^0_1$ and 
$H^{\pm}\rightarrow\tilde\chi^{\pm}_1\nu_{\tau}$ which can be dominant at
low $\tan\beta$ \cite{ChaHiggs}.

\section{BRpV Embedded into Supergravity and the Lightest Higgs Mass}

The BRpV model can be successfully \cite{SugraBRpV} embedded into SUGRA 
with radiative breaking of the electroweak symmetry \cite{RSB} and 
universality of soft masses. The electroweak symmetry is broken 
through the vacuum expectation value (vev) of the tau--sneutrino,
in addition to the two Higgs field vevs, and it contributes to the mass 
of the gauge bosons. The correct vev's are found by imposing the 
three tadpole equations, where one--loop tadpoles corrections are 
important for the tau--sneutrino as well as for the two Higgs fields 
\cite{SugraBRpV}. 

\begin{figure}[htb]
\centerline{\psfig{file=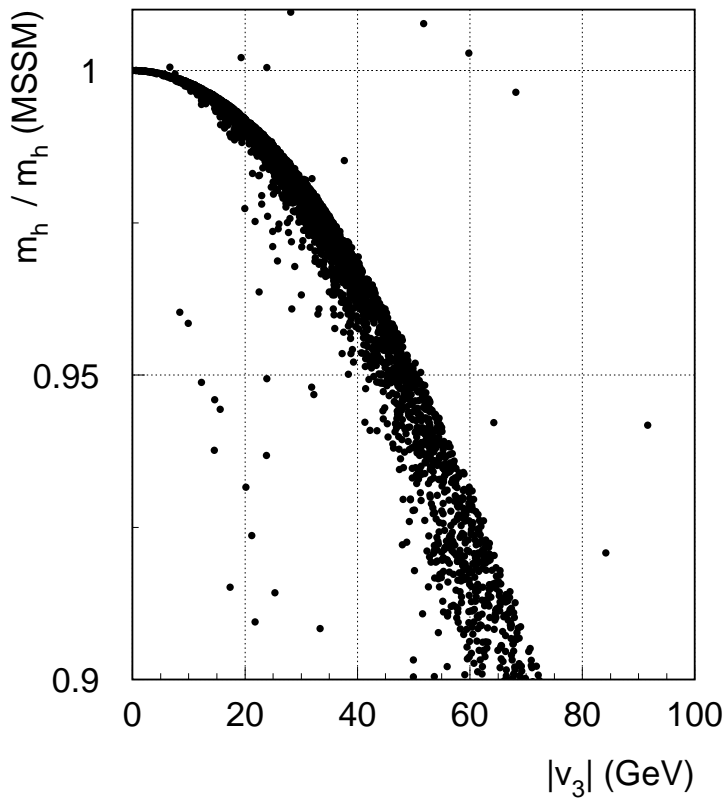,height=11cm}}
\caption{
Ratio between the lightest Higgs boson mass $m_h$ in BRpV and the same 
mass in the MSSM ($v_3=0$ limit) as a function of the tau--sneutrino
vacuum expectation value.
}
\label{ratio_v3}
\end{figure} 
In MSSM--BRpV the CP--even Higgs bosons mix with the real part of the 
tau--sneutrino \cite{CGJRV}. The effect of this mixing is to lower the
mass of the lightest scalar as can be appreciated in Fig.~\ref{ratio_v3},
with the exception of a few, statistically insignificant, exceptional
points. 

\begin{figure}[htb]
\centerline{\psfig{file=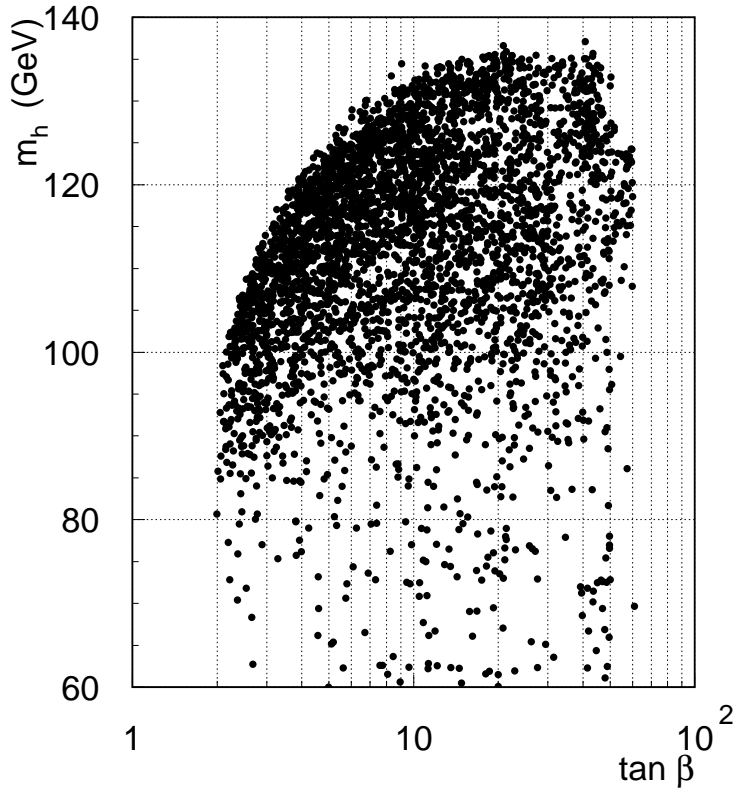,height=11cm}}
\caption{
Lightest neutral Higgs mass in BRpV as a function of $\tan\beta$.
The lower bound on $\tan\beta$ is due to the non--perturbativity of the
top quark Yukawa coupling.
} 
\label{mh_tanbeta_60}
\end{figure} 
In Fig.~\ref{mh_tanbeta_60} we see that the upper bound on the lightest 
CP--even Higgs mass does not change neither. In this figure we have 
included in the $3\times3$ Higgs mass matrix only the radiative corrections
proportional to the fourth power of the top quark mass \cite{DiazHaberii}.
The neglected corrections brings down the upper bound in several GeV. 
As in the MSSM, the upper bound for $m_h$ decreases when $\tan\beta$ 
decreases, reaching in BRpV somewhat less than 100 GeV if $\tan\beta=2$. 
This implies that experimental searches for $h$ at CERN also tests this 
model if $\tan\beta$ is close to unity, as it occurs in the MSSM \footnote{
Simple formulas for the one--loop radiatively corrected lightest neutral
Higgs mass $m_h$ when $\tan\beta$ is close to one can be found in 
\cite{DiazHaberii}.
}. 

It is interesting to note that, since the vacuum stability bound on the SM 
Higgs boson mass is about 135 GeV \cite{Stabil}, the measurement of the Higgs 
boson mass can distinguish between the BRpV model and the SM (with no 
physics below $\sim 10^{10}$ GeV) in the same way as it can distinguish 
between the MSSM (or the NMSSM) and the SM \cite{Vander}.

\section{Unification of Yukawa Couplings}

\begin{figure}[htb]
\centerline{\psfig{file=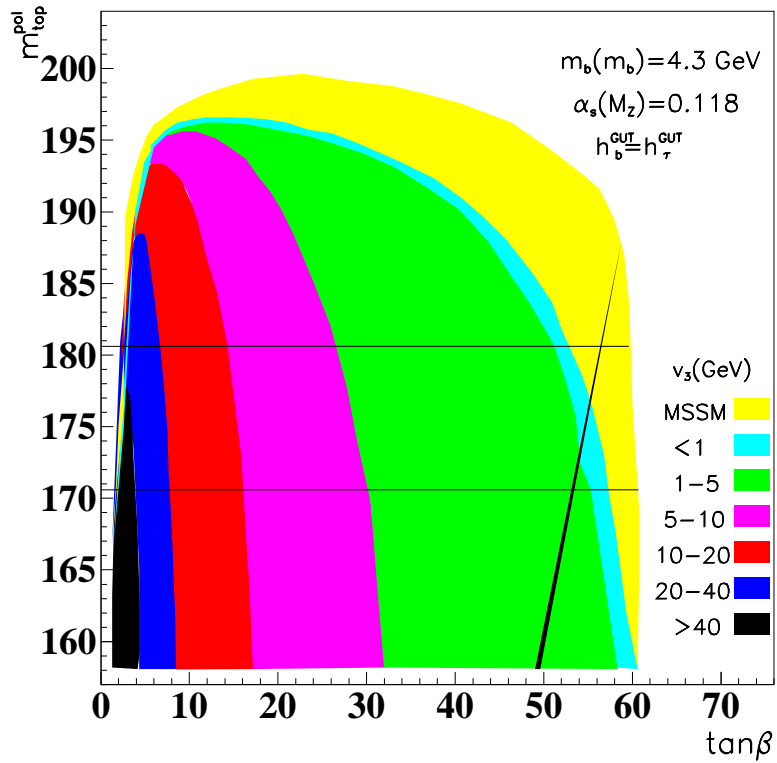,height=11cm}}
\caption{
Regions of the $m_t$-$\tan\beta$ plane where bottom--tau Yukawa coupling
can be achieved in the MSSM and in BRpV. In the case of BRpV regions are 
labeled by the tau--sneutrino vev $v_3$. The inclined straight line 
corresponds to top--bottom--tau Yukawa unification.
} 
\label{aretop}
\end{figure}

In the MSSM, bottom--tau Yukawa unification is achieved at two disconnected
regions at low and high values of $\tan\beta$ \cite{YukUnif}. This can be 
appreciated in the ``inverted U'' shaped region in the plane 
$m_t$--$\tan\beta$ shown in Fig.~\ref{aretop}. The two horizontal lines
correspond to the $1\sigma$ determination of the top quark mass 
\cite{topmass} at Fermilab \footnote{
We note that the new decay modes of the top quark present in the BRpV 
model impose only mild constraints on the BRpV parameter $\epsilon_3$ 
\cite{topBRpV}.
}.

In BRpV the unification of couplings is modified \cite{YukBRpV} in an 
important way (see \cite{AllanachRep} for the TRpV case). In 
Fig.~\ref{aretop} we see that by choosing the value of the tau--sneutrino
vev $v_3$ we can achieve bottom--tau Yukawa unification at any value
of $\tan\beta$ provided we keep inside the perturbativity region
$2\lsim\tan\beta\lsim60$. The high $\tan\beta$ region where top--bottom--tau 
unification is found is twice as big in BRpV compared with the MSSM. The 
$t-b-\tau$ unification is found at values of $v_3<5$ GeV, therefore, it 
would rule out regions of parameter space where the bilinear violation of 
R--Parity is large. We note that in Fig.~\ref{aretop} we define 
$\tan\beta=v_2/v_1$ to preserve the MSSM definition. Another possibility is 
to define $\tan\beta'=v_2/\sqrt{v_1^2+v_3^2}$ which has the advantage of 
being invariant under rotations on the $L_3-H_1$ plane. We have checked 
that Fig.~\ref{aretop} does not change appreciably when plotted against 
$\tan\beta'$.

The reason why $b-\tau$ unification in BRpV fills the intermediate regions 
of $\tan\beta$ can be understood as follows. First of all, we notice that
the quark and lepton masses are related to the different vevs and Yukawa 
couplings in the following way
\begin{equation}
m_t^2=\half h_t^2v_2^2\,,\qquad m_b^2=\half h_b^2v_1^2\,,\qquad
m_{\tau}^2=\half h_{\tau}^2v_1^2(1+\delta)\,,
\label{fermmass}
\end{equation}
where $\delta$ depends on the parameters of the chargino/tau mass matrix
and is positive \cite{ChaHiggs,YukBRpV}. This implies that the ratio of 
the bottom and tau Yukawa couplings at the weak scale is given by
\begin{equation}
{{h_b}\over{h_{\tau}}}(m_{weak})={{m_b}\over{m_{\tau}}}\sqrt{1+\delta}
\label{ratio1}
\end{equation}
and grows as $|v_3|$ is increased.

On the other hand, if $h_b$ and $h_{\tau}$ unify at the GUT scale, then 
at the weak scale its ratio can be approximated by
\begin{equation}
{{h_b}\over{h_{\tau}}}(m_{weak})\approx exp\left[
{1\over{16\pi^2}}\left({16\over3}g_s^2-3h_b^2-h_t^2\right)
\ln{{M_{GUT}}\over{m_{weak}}}\right]
\label{hb_htau}
\end{equation}
implying that the combination $3h_b^2+h_t^2$ should decrease when $|v_3|$
increases.

In the MSSM region of high $\tan\beta$ the bottom quark Yukawa coupling
dominates over the top one, and the opposite happens in the region of
low $\tan\beta$. Therefore, at high (low) values of $\tan\beta$, the
Yukawa coupling $h_b$ ($h_t$) will decrease if $|v_3|$ increases, which
implies an increase of $v_1$ ($v_2$) in order to keep constant the quark
masses. Similarly, in order to keep constant the $W$ mass,
$m_W^2=\fourth g^2(v_1^2+v_2^2+v_3^2)$, the vev $v_2$ ($v_1$) decreases
at the same time. This implies that unification occur at lower (higher)
values of $\tan\beta$ as $|v_3|$ increases, explaining what we see in 
Fig~\ref{aretop}.

The fact that bottom--tau unification occurs at any value of $\tan\beta$
in BRpV seems important to us considering that in the MSSM the low 
$\tan\beta$ region is disfavoured by the non observation of the lightest 
Higgs boson. In addition, the high $\tan\beta$ region is disfavoured
because it is usually difficult to find the correct electroweak symmetry 
breaking. These difficulties are avoided in BRpV if the sneutrino vev is
sufficiently large.

\section{Conclusions}

In its simplest form, Bilinear R--Parity Violation is a one parameter
extension of the MSSM which can be successfully embedded into SUGRA with 
radiative breaking of the electroweak symmetry and universality of soft
masses. This is achieved through the running of the same RGEs of the
MSSM since BRpV does not introduce new interactions. Therefore it is a 
very simple framework to study R--Parity violating phenomena. In addition,
BRpV generates a tau--neutrino mass which, in models with universality of
soft masses, is radiatively generated and proportional to the bottom
quark Yukawa coupling squared, therefore, naturally small.

In BRpV charged Higgs bosons mix with the staus, and because of this,
staus contribute to the decay $b\rightarrow s\gamma$. In an unconstrained
version of the model we have showed that the bounds on the charged Higgs
boson mass from $B(b\rightarrow s\gamma)$ are relaxed by $\sim 100$ GeV
in the heavy squark limit, and by $\sim 30$ GeV in the light chargino 
and light charged Higgs limit. In this case, charged Higgs lighter that
the $W$--gauge boson are possible and observable at LEP2. 
Nevertheless, R--Parity violating decay modes will compete with
the traditional decay modes of the charged Higgs in the MSSM. In a 
similar way, the neutral CP--even Higgs bosons mix with the real part
of the tau--sneutrino. In general, this mixing lowers the Higgs mass but
leaves the upper bound unchanged.

Finally, we have shown that it is much easier to find unification of the
bottom and tau Yukawa couplings in BRpV than in the MSSM. By choosing 
the value of the tau--sneutrino vacuum expectation value, $b-\tau$
unification can be achieved at any value of $\tan\beta$. Unification
of $t-b-\tau$ can be found at high values of $\tan\beta$ but in a region
twice as large than in the MSSM.

\section*{Acknowledgements}

I am thankful to my collaborators A. Akeroyd, J. Ferrandis, 
M. Garcia--Jare\~no, W. Porod, D. Restrepo, J. Rom\~ao, 
E. Torrente--Lujan, and J. Valle for their contribution to the work 
presented here. This research has been supported in part by the
U.S. Department of Energy under contract number DE-FG02-97ER41022
and in part by a grant from the Ministerio de Educaci\'on y Ciencias
in Spain.

\end{document}